\documentclass{acm_proc_article-sp}

\usepackage{graphicx}
\usepackage{epsfig}

\usepackage{color}
\usepackage[english]{babel}
\usepackage[utf8]{inputenc}
\usepackage[colorinlistoftodos]{todonotes}

\usepackage{amsmath}    
\usepackage{mathtools}  
\usepackage{amssymb}    
\usepackage{verbatim}   
\usepackage{color}      
\usepackage{hyperref}   
\usepackage{multicol}  %
\usepackage{multirow}
\usepackage{ifthen}  
\usepackage{float}

\usepackage{graphicx}
\usepackage{caption}
\usepackage{subcaption}

\usepackage{lipsum}       
\usepackage{changepage}   
\usepackage{enumitem}
\usepackage{flushend}

\begin{document}

\conferenceinfo{}{Bloomberg Data for Good Exchange 2016, NY, USA}

\title{Data Science in Service of Performing Arts: Applying Machine Learning to Predicting Audience Preferences}

\numberofauthors{510}
\author{
\alignauthor Jacob Abernethy\\
       \affaddr{University of Michigan}\\
       \affaddr{Ann Arbor, MI}\\
       \email{jabernet@umich.edu}
\alignauthor Cyrus Anderson\\
       \affaddr{University of Michigan}\\
       \affaddr{Ann Arbor, MI}\\
       \email{andersct@umich.edu}
\alignauthor Alex Chojnacki\\
       \affaddr{University of Michigan}\\
       \affaddr{Ann Arbor, MI}\\
       \email{thealex@umich.edu}
\and
\alignauthor Chengyu Dai\\
       \affaddr{University of Michigan}\\
       \affaddr{Ann Arbor, MI}\\
       \email{daich@umich.edu}
\alignauthor John Dryden\\
       \affaddr{University of Michigan}\\
       \affaddr{Ann Arbor, MI}\\
       \email{jcdryden@umich.edu}
\alignauthor Eric Schwartz\\
       \affaddr{University of Michigan}\\
       \affaddr{Ann Arbor, MI}\\
       \email{ericmsch@umich.edu}
\and
\alignauthor Wenbo Shen\\
       \affaddr{University of Michigan}\\
       \affaddr{Ann Arbor, MI}\\
       \email{shenwb@umich.edu}
\alignauthor Jonathan Stroud\\
       \affaddr{University of Michigan}\\
       \affaddr{Ann Arbor, MI}\\
       \email{stroud@umich.edu}
\alignauthor Laura Wendlandt\\
       \affaddr{University of Michigan}\\
       \affaddr{Ann Arbor, MI}\\
       \email{wenlaura@umich.edu}
\and
\alignauthor Sheng Yang\\
       \affaddr{University of Michigan}\\
       \affaddr{Ann Arbor, MI}\\
       \email{physheng@umich.edu}
\alignauthor Daniel Zhang\\
       \affaddr{University of Michigan}\\
       \affaddr{Ann Arbor, MI}\\
       \email{dtzhang@umich.edu}
}
\maketitle

\begin{abstract}
Performing arts organizations aim to enrich their communities through the arts. To do this, they strive to match their performance offerings to the taste of those communities. Success relies on understanding audience preference and predicting their behavior. Similar to most e-commerce or digital entertainment firms, arts presenters need to recommend the right performance to the right customer at the right time. As part of the Michigan Data Science Team (MDST), we partnered with the University Musical Society (UMS), a non-profit performing arts presenter housed in the University of Michigan, Ann Arbor. We are providing UMS with analysis and business intelligence, utilizing historical individual-level sales data. We built a recommendation system based on collaborative filtering, gaining insights into the artistic preferences of customers, along with the similarities between performances. To better understand audience behavior, we used statistical methods from customer-base analysis. We characterized customer heterogeneity via segmentation, and we modeled customer cohorts to understand and predict ticket purchasing patterns. Finally, we combined statistical modeling with natural language processing (NLP) to explore the impact of wording in program descriptions. These ongoing efforts provide a platform to launch targeted marketing campaigns, helping UMS carry out its mission by allocating its resources more efficiently. Celebrating its 138th season, UMS is a 2014 recipient of the National Medal of Arts, and it continues to enrich communities by connecting world-renowned artists with diverse audiences, especially students in their formative years. We aim to contribute to that mission through data science and customer analytics.
\end{abstract}


\section{Introduction}
The University Musical Society (UMS) is a 501(c)(3) non-profit performing arts organization affiliated with the University of Michigan in Ann Arbor. UMS seeks to engage the community with new and innovative artists. Founded in the winter of 1880, UMS is one of the oldest non-profit performing arts presenters in the country and presents 65 to 75 shows per year in a variety of genres. UMS holds events in the Ann Arbor area across multiple venues. The largest is Hill Auditorium, which has a maximum capacity of 3,536 and accommodates world-class musicians and performers.

On the one hand, UMS is a non-profit organization, depending on donations and grants for funding. On the other, it depends on generating revenue through ticket sales. Much like many other businesses, UMS uses common tactics in multi-channel retail, online marketing, and digital content distribution to maximize its marketing strategy. UMS maintains a large purchase history database, containing information such as what marketing material was sent to customers, how tickets were purchased, and how users' purchasing habits have changed over time.

In this paper, we explore how this rich dataset can be further utilized to gain a better understanding of the UMS audience to eventually guide marketing and event programming decisions. We show how machine learning and other data analysis techniques can be applied to gain insight into purchasing patterns, as well as suggest improvements in future marketing strategy. Section 2 positions this work in the context of other research. Section 3 describes the main UMS dataset, as well as additional data collected, and provides model-free visualizations of the data patterns. Section 4 examines the performances themselves, specifically through the language used in descriptions. Section 5 uses collaborative filtering techniques to analyze purchase history data. Section 6 shows how Markov chains can be used to model the data, linking to a broader stream of statistical modeling in customer-base analysis. Finally, we end in Section 7 with conclusions and future work.

\section{Related Work}
Prior to this, performing arts organizations have used data analytics to better understand their customer bases. There has been work done on segmenting customers into well-defined groups based on their purchasing behaviors \cite{brown2007segmentation,mitchell1984professional}. Our work leverages new advances in machine learning and data science to better understand customer behavior.

One of the techniques that we incorporate is text analysis. We use stylistic features of the text, which have previously been used to predict things like document authorship and genre \cite{de2000mining,karlgren1999stylistic}. Another contribution is that we demonstrate how to utilize recent developments in matrix factorization \cite{koren2009matrix} to understand customer artistic preferences and how ticket sales are correlated based on the artistic styles of performances. In our method, all of these analytics can be extracted from easily accessible ticket purchase data, without organizing dedicated customer surveys. Finally, we use customer lifecycle analysis, modeling customers' ability to buy with Markov chains \cite{chung1967markov}, similar to \cite{morrison1966testing}, where Markov chains were used to model customers' switching between different brands.

\section{Data}
UMS has provided us with five years of anonymized ticket purchasing data from 2011 to 2015. We use the first three years for training and hold out the most recent two years for test data. This data set includes over 190,000 transactions from 48,000 users, totaling over \$13 million in revenue. Each transaction contains the following pieces of information:
\begin{itemize}
	\item UMS account number
    \item Date the account was created (or digitalized, for pre-digital accounts)
    \item Customer type (either a household, individual, or organization)
    \item Name, date, and venue of the performance
    \item Price of the tickets and number of seats sold
    \item Information about whether the ticket was part of a promotion or special offer
    \item Mode of sale (how the ticket was bought, for example, via the UMS website or over the phone)
    \item Date of the order
    \item Postal code of the customer
\end{itemize}

One particularly important dynamic captured in the dataset is information about subscriptions. At the beginning of each performance season, UMS offers customers the chance to purchase subscriptions, which are packages of tickets for a series of shows. Usually each subscription is thematic and includes related shows. Some current subscription series are Dance and Theater, Jazz, and Choral / Vocal. Subscriptions often need to be treated differently during data analysis because when customers buy a series, they do not individually select each show. This changes their purchasing patterns and behaviors.

\subsection{Data Visualizations}

    \begin{figure*}[t!]
    	\centering
    	\includegraphics[width=\textwidth]{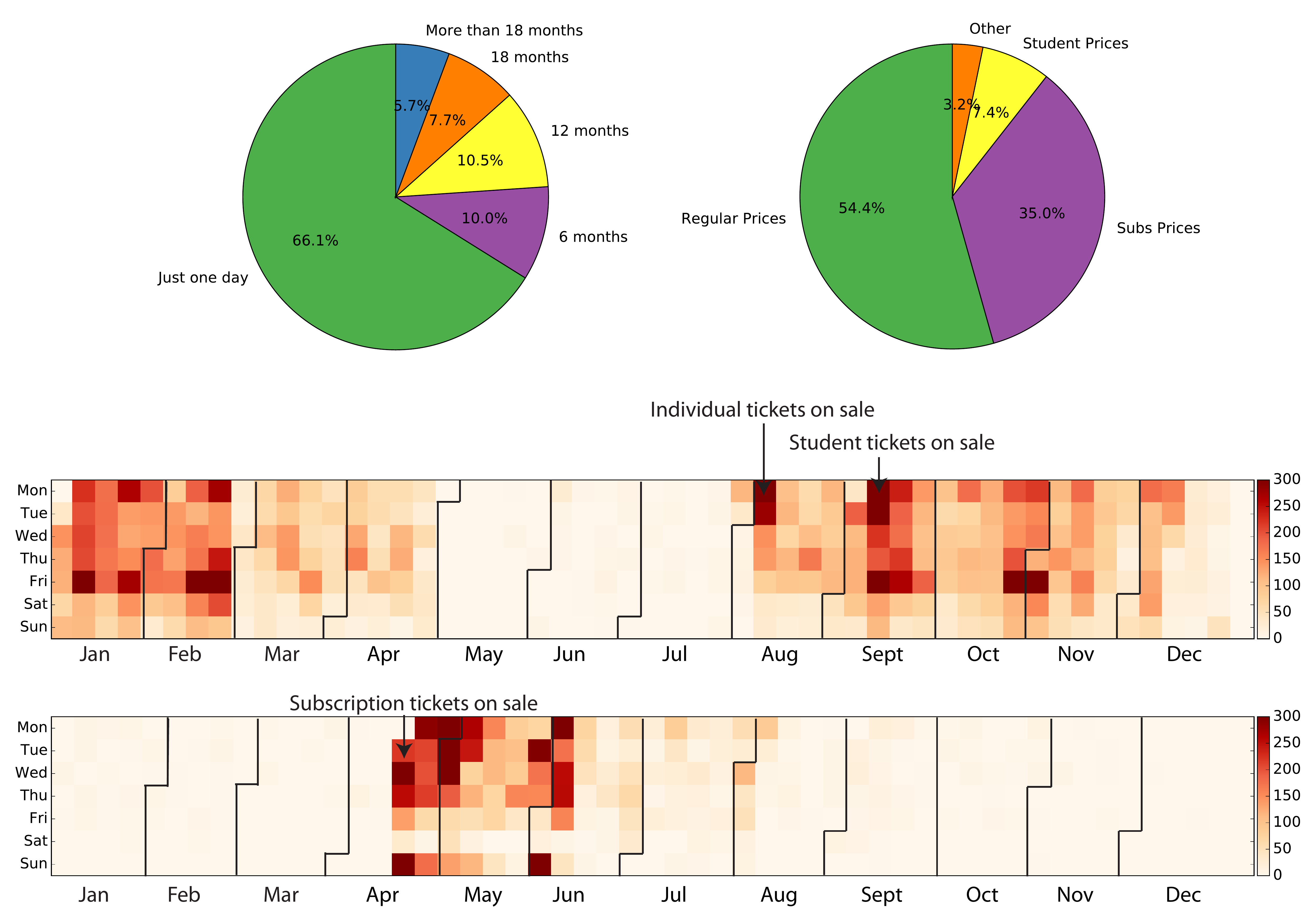}

    	\caption{Top left pie chart: the duration of customers' activities. The duration is defined as the time span between the customer's first purchase and his or her last purchase. Top right pie chart: UMS revenue composition by the price groups: regular price, subscription price, student price, or other. Top heat map: the number of non-subscription tickets bought throughout the year. Darker colors indicate that more tickets were bought on that day. Bottom heat map: the number of subscription tickets bought throughout the year. For both heatmaps, purchase data from 2013 is used.}
    	\label{viz}
    \end{figure*}
	Data visualizations reveal patterns in terms of customer activity, revenue composition, and purchase patterns. Figure \ref{viz} shows select statistics and visualizations from the UMS dataset. We define the activity duration of each customer as the time span between the first purchase and the last purchase. Interestingly, 66\% percent of customers made only one purchase, never returning to buy another ticket. This suggests that there is great potential to convert these one-time customers into frequent customers. The importance of frequent customers is highlighted in the pie chart of the revenue composition. This pie chart shows that more than one-third of all revenue is from subscription purchases. This relatively large percentage of revenue is from a relatively small fraction of customers, as only 5.6\% customers are subscription buyers. High purchase concentration is to be expected with a large heterogeneous audience. Finally, we find that the non-subscription purchases and subscription purchases have distinct time patterns. Non-subscription (regular ticket) purchases are distributed throughout the year. There are almost no non-subscription purchases in the months of May, June, and July since there are generally no performances scheduled during these months. In addition, at the opening days of various ticket groups, there is always a rush for tickets. Contrary to the non-subscription purchases, subscription purchases are concentrated from April to June when there are very few performances. This is understandable since the subscriptions are for the coming season from September to April of next year. The rushes for tickets in both patterns imply that performances organized by UMS are very popular and well-received. 
    
\subsection{Performance Descriptions}
\begin{figure*}[t!]
\centering
\includegraphics[width=\textwidth]{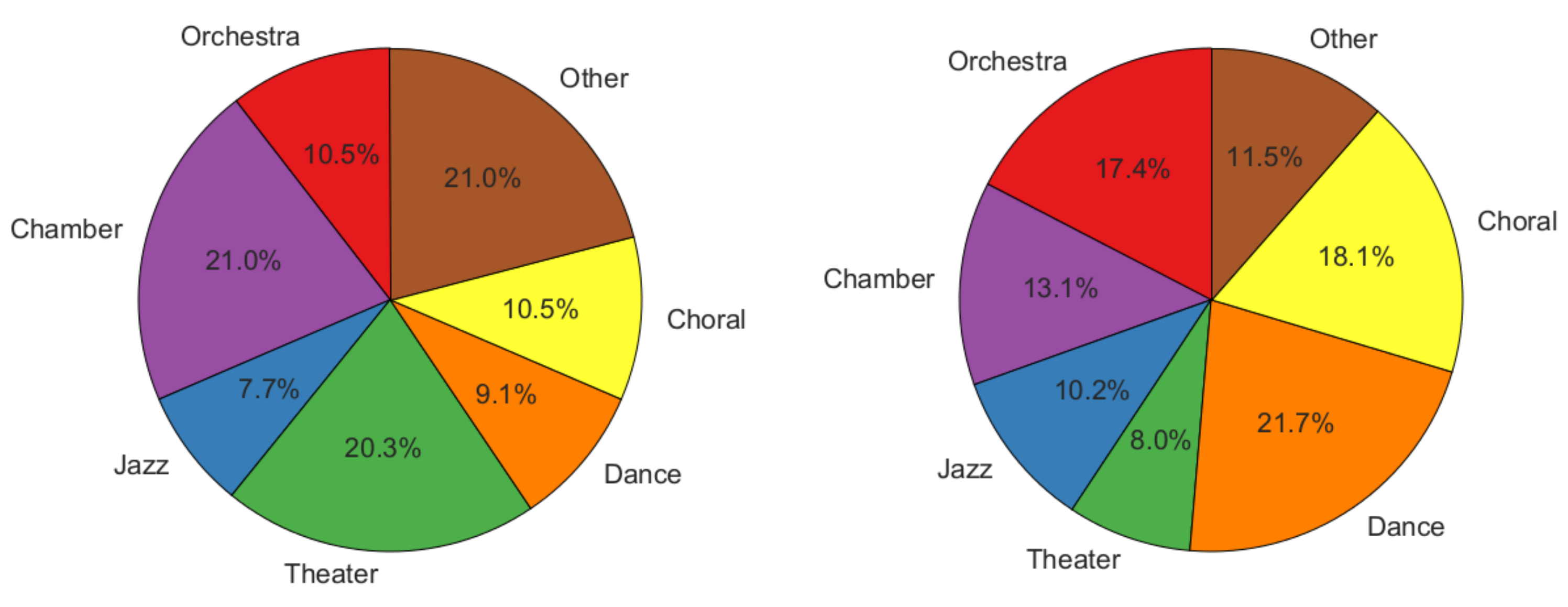}
\caption{The left pie chart shows performance descriptions broken down by category. The right pie chart compares the total number of seats sold in each genre.}
\label{category_pie}
\end{figure*}

To augment the purchasing data provided by UMS, we collect descriptions for each performance. These descriptions are written by UMS and are publicly available on the UMS website. The average description length is 164.4 tokens, where each token is either a word or a symbol. After the descriptions are collected, they are manually categorized into seven categories: Orchestra, Chamber, Jazz, Theater, Dance, Choral, and Other. Figure \ref{category_pie} shows the percentage of performance descriptions in each category, as well as the total number of seats sold in each genre. These charts highlight some discrepancies in the number of shows offered for each genre and the popularity of each genre. For example, 9.1\% of performances are in the Dance category, while 21.7\% of the total number of seats sold are to Dance shows.

\section{Wording in Performance Descriptions} \label{sec:nlp}
One of the marketing tools that UMS has at their disposal is the written performance descriptions distributed via programs, brochures, posters, and online media. Analyzing the writing style of these descriptions can provide insight into why customers choose to see the shows that they do. They also help to explain similarity of performances and artists.

Several metrics exist to measure the style of a piece of writing. One of these is readability, which assigns a reading grade level to a piece of writing. One standard measure of readability is the Flesch-Kincaid Grade Level \cite{kincaid1975derivation}. This measure is often used to measure the complexity of a piece of literature \cite{agichtein2008finding,friedman2006systematic}. The grade level is calculated according to the following formula:
\begin{equation}
0.39\frac{\mbox{total words}}{\mbox{total sentences}} + 11.8\frac{\mbox{total syllables}}{\mbox{total words}} - 15.59
\end{equation}

Another text-based style metric is formality. Formality attempts to quantify the preciseness and informativeness of a statement. The Heylighen and Dewaele measure of formality is calculated according to the following formula \cite{heylighen1999formality}:
\begin{equation}
\begin{split}
(\mbox{noun freq.} + \mbox{adjective freq.} + \mbox{preposition freq.} + \\ \mbox{article freq.} - \mbox{pronoun freq.} - \mbox{verb freq.} - \\ \mbox{adverb freq.} - \mbox{interjection freq.} + 100)/2
\end{split}
\end{equation}

\begin{figure*}[t!]
\centering
\includegraphics[width=\textwidth]{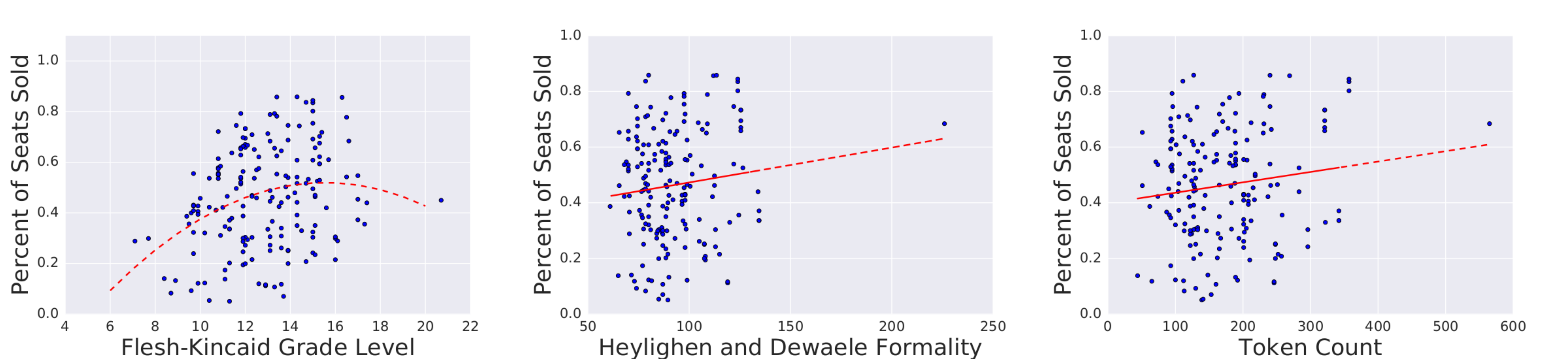}
\caption{Style metrics applied to the written performance descriptions. These graphs show the relationship between three style metrics (readability, formality, and length) and the percentage of seats sold for that show. These plots do not include tickets sold as part of a subscription. Each graph is fitted with either a 2-dimensional or a 1-dimensional line, to show the trends in the data. The Pearson correlation coefficients for these scatter plots are, from left to right, 0.26, 0.12, and 0.14.}
\label{style_metrics}
\end{figure*}

A third text-based style metric is the length of a document. Figure \ref{style_metrics} shows the relationship between these three metrics and the percentage of seats sold for every show. Only tickets that are not part of a subscription are included for this analysis. A 2-dimensional line of best fit is fitted to the readability plot, while 1-dimensional lines of best fit are fitted to the other plots. These plots show that on average, as formality and description length increase, there is an increase in the number of tickets sold for that performance. There is a similar trend for readability, but this data is better fitted with a polynomial curve, indicating that the optimal readability for a program description is around grade level 15.


\section{Collaborative Filtering Modeling for Purchase History}
In addition to analyzing performance descriptions, another way to understand historical purchasing patterns is through the use of a model based on collaborative filtering. \cite{bell2007lessons,koren2009matrix}. Popularized by the Netflix Prize in 2006, collaborative filtering is a technique that automatically matches customers to performances they might enjoy based on information such as purchase history and customer or content similarity.  Unlike Netflix, where content remains available to all users, a performing arts organization has live performance constraints that come and go. Because of this, the collaborative filtering approach applied will have to make recommendations among a small set of possible shows remaining in the given season. We return to this as future work in the final section.

Another interesting component of this collaborative filtering modeling problem is the diversity of performances across genres. Due to the great variety of UMS performances, it is difficult to define a reliable similarity metric that can be used to compare performances. Therefore, our system relies primarily on well-documented recent purchase history. Applying matrix factorization (MF) to collaborative filtering has achieved great success in both academic research and industrial application. We adopt this approach, which is formally introduced below. 
        \begin{figure*}[t!]
          \begin{minipage}[t]{0.46\linewidth}
            \centering
            \includegraphics[width=\textwidth]{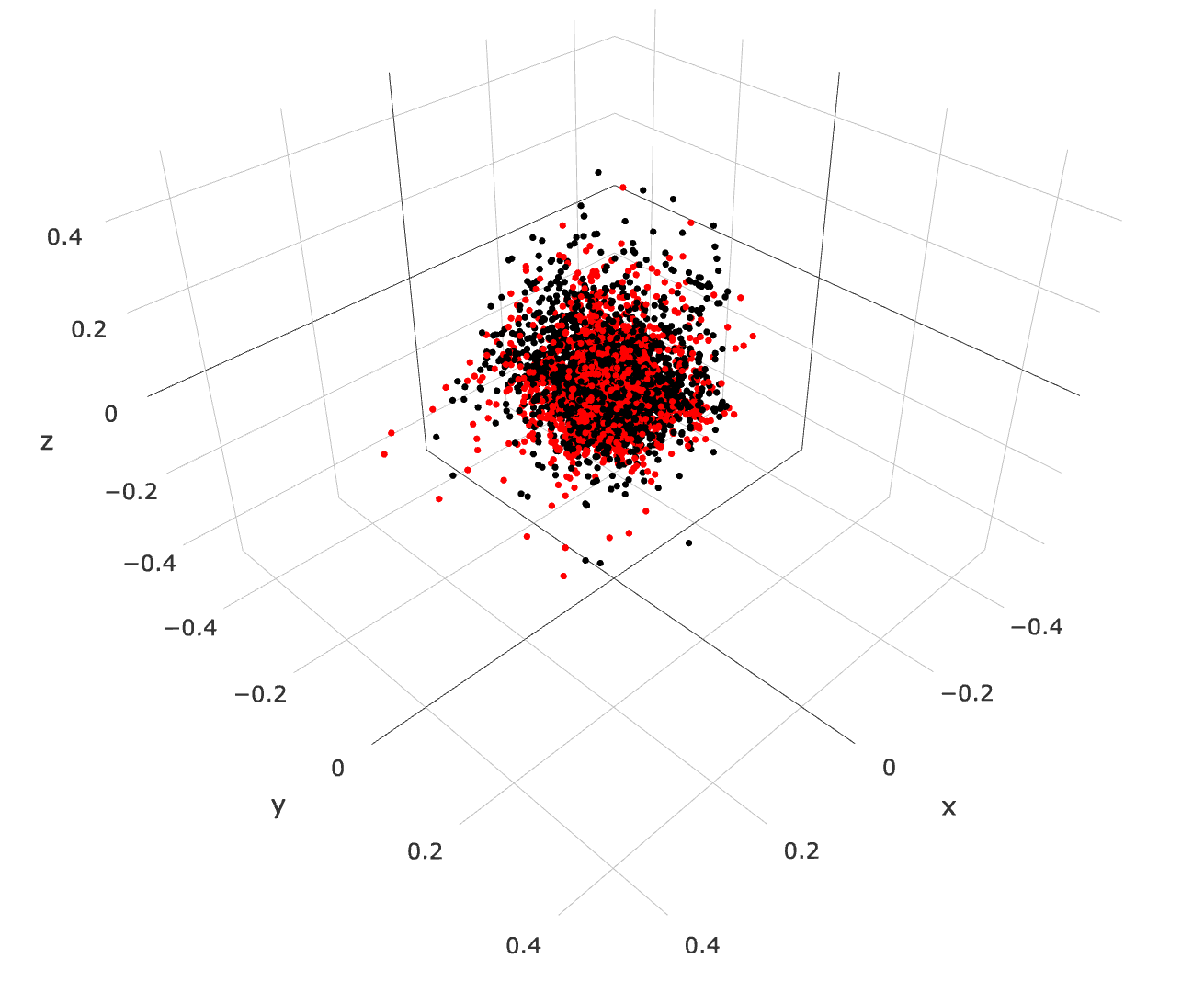}
            \caption{Visualizing artistic preferences of both students (red) and the general public (black). This scatter plot shows the embeddings of different customers in a latent space representing artistic style preferences.}
            \label{lat_cust}
          \end{minipage}
          \hspace{0.04\linewidth}
          \begin{minipage}[t]{0.46\linewidth}
            \centering
			\includegraphics[width=\textwidth]{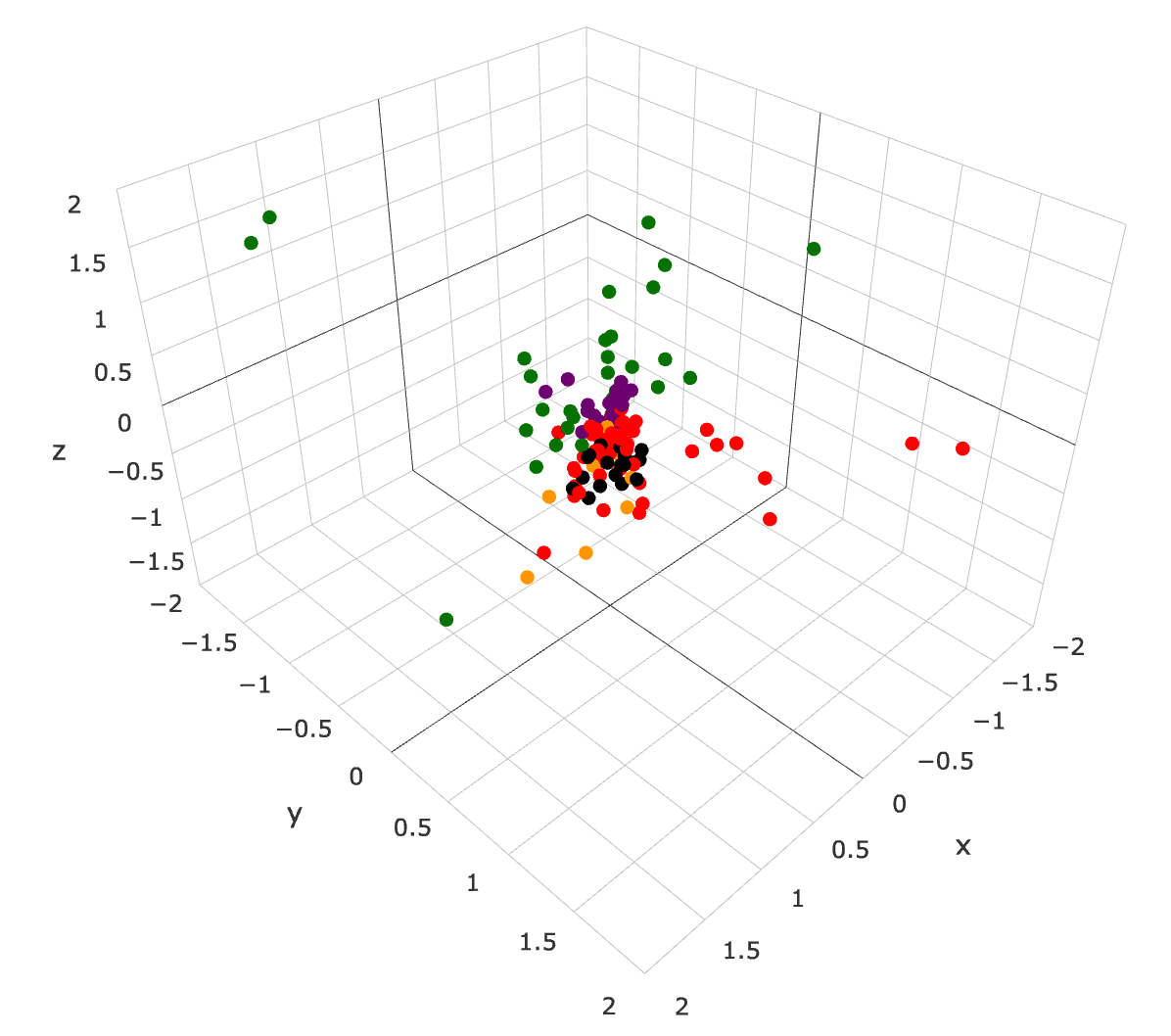}
			\caption{Visualizing artistic styles for performances that belong to different subscription series. Green: Choral Unions, Purple: Chamber Music, Black: Dance, Orange: Jazz, Red: Others}
            \label{lat_perf}
          \end{minipage}
        \end{figure*} 
	\subsection{Methodology}
        First, we represent purchase history as a binary-valued matrix $\mathbf{X}$ with dimensions $N_c$ x $N_p$, where $N_c$ and $N_p$ are the numbers of unique customers and performances, respectively, in the dataset. $\mathbf{X}_{ij} = 1$ indicates that customer $i$ purchased a ticket for performance $j$ , and $\mathbf{X}_{ij} = 0$ indicates that no ticket was purchased.
        
Every customer's willingness to purchase a ticket for a specific performance can be expressed in terms of how well that performance matches each of the artistic styles the customer is interested in. We also include constant bias terms to model that some customers have more buying power and similarly some performances are more popular than others. This gives us the following equation.
        
            \begin{equation}
				\mathbf{X}_{ij} \approx \sum_{l}\mathbf{L}_{il}\mathbf{R}_{jl} + \mathbf{B}^{L}_{i} + \mathbf{B}^{R}_{j}
			\end{equation}
        
        Translating into the language of matrices,
            
            \begin{equation}
				\mathbf{X} \approx \mathbf{L}\mathbf{R}^T + \mathbf{B}^{L} + \mathbf{B}^{R}
			\end{equation}
            
            A more formal way of posing the problem is to express it in terms of a regularized Frobenius norm optimization problem, as follows:
			\begin{equation*}
                \begin{aligned}
                & \underset{\mathbf{L}, \mathbf{R}, \mathbf{B}^{L}, \mathbf{B}^{R}}{\text{minimize}}
                & &\|\mathbf{X} - \mathbf{L}\mathbf{R}^T - \mathbf{B}^{L} - \mathbf{B}^{R}\|^2_{F} + \frac{\lambda}{2}\|\mathbf{L}\|_F + \frac{\lambda}{2}\|\mathbf{R}\|_F \\ 
                & \text{subject to}
                & & \mathbf{L} \in \mathbb{M}_{N_c,L} , \mathbf{R} \in \mathbb{M}_{N_p,L}, \mathbf{B}^{L}, \mathbf{B}^{R}, \in \mathbb{M}_{N_c,N_p} 
                \end{aligned}
            \end{equation*}
            
            In the above equation, $\mathbf{B}^{L}$ / $\mathbf{B}^{R}$ are column-wise / row-wise constant matrices respectively. 
            
This optimization problem is closely related to singular value decomposition (SVD). It is generalized to include a constant term in the formula and to make training possible even when some of the data is not present.
            
            Standard approaches to training this model include stochastic gradient descent (SGD) and alternating least square (ALS). We adopt the latter approach, because empirically we found that SGD with all random initialization fails to converge when the matrix is tall-and-skinny (when the number of customers greatly exceeds the number of performances). Note that if we fix either $\mathbf{L}$ or $\mathbf{R}$ and optimize with respect to the other, then the problem becomes a standard quadratic matrix optimization problem which can be solved by least squares. Alternating between fixing either $\mathbf{L}$ or $\mathbf{R}$ and solving for the other, the algorithm converges reasonably well after a few hundred steps for practical use and further study.

  	\subsection{Interpreting the Factorization}
This simple matrix factorization model can provide insight into customers and performances, based solely on the purchase history. $\mathbf{L}$ and $\mathbf{R}$ have clear geometric interpretations as collections of vectors of each customer or performance's position in the latent space of artistic style. For each customer vector, the magnitude of the vector is related to the total purchases of the customer, while the direction of the vector represents the customer's artistic style preferences. For each performance vector, the magnitude of the vector is related to the number of seats sold for the performance, while the direction of the vector represents the artistic style of the performance. Each point in the latent space represents a vector connecting the origin to that point.
        
        \subsubsection{Customer Preference Analysis}
These customer vectors can be used to explore customer heterogeneity in taste. Since UMS is affiliated with the University of Michigan, we will use those vectors to compare university students' artistic preferences with the preferences of the non-student general public. This analysis furthers one of the core goals of UMS, which is to enrich students' cultural experiences. To investigate this question, we restrict our model to the three most significant latent dimensions, to allow for easy visualization. We separate out students by looking for customers who have purchased tickets from a student promotion or at a student price.

        From the visualization of Figure \ref{lat_cust}, we conclude that both students and the general public have large in-group variation of artistic style preferences. This is reflected in the diverse directions of the latent vectors; they cannot be simply clustered into a few definite groups. The great spectrum of performance types that UMS currently provides serves this diverse community well. 
        
        More importantly, our result also suggests that in general students do not have different artistic preferences than regular customers. However, the latent vectors representing student customers have smaller magnitudes, meaning that students in general have less willingness to purchase. This certainly has face validity, as students typically have smaller incomes than working adults. This analysis, while not able to make strong causal claims, suggests that UMS may be wise to continue their current student discount pricing policy, which is giving general discounts to students while not limiting the discounts to any specific types of performances.

    	\subsubsection{Performance Style Analysis}
   		Another question of interest is if performances in the same subscription series are also similar in terms of artistic style. We investigate this question by specifically analyzing the purchase data excluding subscription tickets. Performance vectors with similar directions indicate similar artistic styles, as shown in Figure \ref{lat_perf}. In general, Jazz (orange) performances and Choral Union (green) series do show similarity with others of their kind. Chamber (purple) and Dance (black) performances are represented by latent vectors with very small magnitudes due to limited venue sizes relative to other genres. Additionally, a few Other (red) performances approximately form a straight line, showing very related style. Many of these are annually recurring performances of Handel’s Messiah, an Ann Arbor traditional holiday show that many people attend.
 In general, we found individual ticket purchase patterns match the hand-picked subscription series based on genre, as expected. 

    \subsection{Discussion on Missing Data}
As most of UMS's customers come from the local college town of Ann Arbor, many of them have only been residents of the town for a few years. This must be taken into account when considering a customer's willingness to purchase a ticket. If a person is not living in Ann Arbor at the time of the performance, he or she should not be considered unwilling to purchase the ticket. Instead, that information should be considered missing. A reasonable approximation of the customer's arrival time is the customer's UMS account creation date. 

Our training method ALS supports the use of missing entries, as long as we solve for each $\mathbf{L}_i$ vector sequentially, by doing regression only on the performance entries $j$ that are not missing for a particular customer (likewise for training $\mathbf{R}_j$). After this correction, the model contains significantly fewer artifacts and is more interpretable. 
        
\section{Customer Lifecycle Analysis}

\begin{figure*}[t!]
\centering
\includegraphics[width=\textwidth]{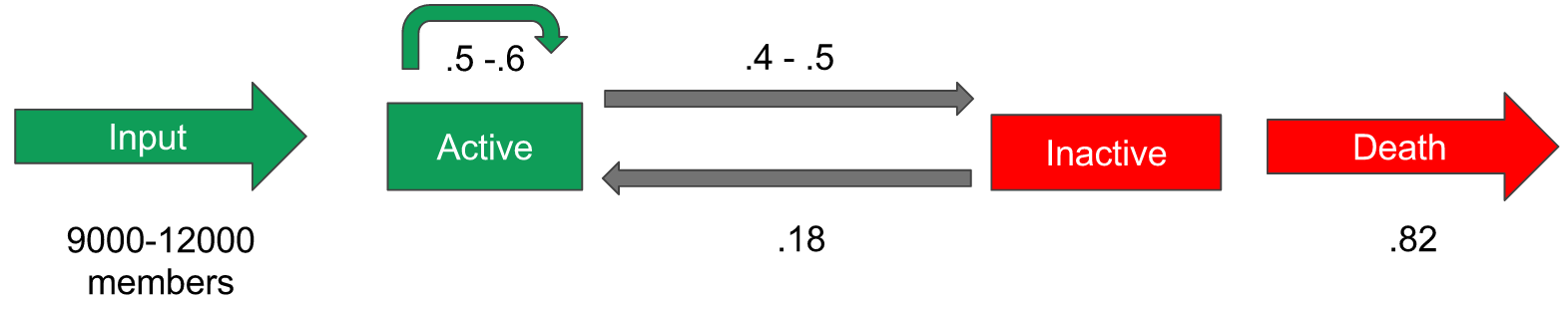}
\caption{A Markov chain showing the lifecycles of UMS customers. The three stages in this cycle are ``active", ``inactive", and ``death". The probabilities of moving from one stage to another are shown in the diagram. The transitions occur annually.}
\label{markov}
\end{figure*}

Collaborative filtering provides a way to analyze the artistic preferences of users. Another way to understand transaction history is to analyze dynamics of purchase behavior throughout a customer's lifecycle, with an aim towards calculating customer lifetime value. This information is particularly helpful for identifying which customers will be most valuable in the future (and which should be ignored). Knowing this can improve the efficiency of targeted promotional efforts, from costly postal mail catalogs to more personalized emails.

We draw on a rich literature of customer-base analysis and customer relationship management in marketing \cite{faderhardie2009}. To begin, we will use a lifecycle consisting of three stages: `active', `inactive', and `dead'. While these can be inferred from the data as latent states, we provide an analysis based on observed stages. If a customer purchases any tickets in a year, he or she is associated with the active state. The inactive state means the customer purchases no tickets in a given year, and the dead state is reached if a customer is inactive for two or more years. These three lifecycle states can be modeled using a Markov chain. A Markov chain models system dynamics as a set of states, where at the end of each season, a customer can transition from one state to any other state according to a set of probabilities. The transition probabilities depend only on the current state. Fitting the model then entails finding the nonzero transition probabilities.

The fitted Markov chain is shown in Figure \ref{markov}. Nearly half the customers in a given cohort become inactive, and of the inactive customers, most do not return, i.e., are churned. We see that UMS sustains this turnover with a number of completely new customers each year. Nevertheless, understanding customer churn is important in balancing  customer acquisition and retention efforts.


\section{Conclusion \& Future Work}

The work presented in this paper analyzes historical purchase history data for UMS. We analyze the correlation between program description wording and ticket sales and utilize collaborative filtering techniques to understand customer artistic preferences and performance styles. We also model the lifecycle of customers as a Markov chain, showing how customer activity falls into regular patterns.

This work is ongoing in all three fronts. The Markov model of customer lifecycles presented here relies on `observed' states with predefined labels. But future work will infer `unobserved' states using hidden Markov models. A special case of these models is known as `Buy till you die' models \cite{faderhardie2009}. By allowing for customer churn to be inferred, we will aim for deeper understanding of purchase behavior throughout the customer lifecycle.

Moving forward, we are working on building an intelligent recommendation system that will be able to predict which customers will enjoy new shows. While this builds on collaborative filtering techniques, it is made more difficult by the `cold start problem.' This occurs when there is no available purchasing information for previously unseen shows.

One solution to this problem is to compare new shows with previous shows that are similar in content. This requires having some means of assessing similarity between shows. We have begun working on a way to do this using topic modeling on performance descriptions. Topic models extract a set of topics from a corpus of training documents. Each topic is a distribution of words, and it describes which words are used most frequently in that particular topic. Then, each document is assigned a distribution of topics. Documents that have similar topic distributions should have similar content. Topic modeling can be used to assign topic distributions to each performance description and find similar performances.

In order to have an effective topic model, it must be trained over a large corpus of documents. The performance descriptions are short and do not provide enough information to train a good model. We have begun collecting a much larger corpus of performance-related documents to train a model. This larger corpus consists of Wikipedia pages that are relevant to different performance genres, such as theater, dance, and orchestra. The larger corpus will be used to train an effective topic model, which will then be used to find similarity scores between different performance descriptions. This information will be input to the collaborative filtering algorithm to complete a fully functional recommendation system. 

From a methodological perspective, we see a promising direction in combining these three areas: topic modeling,  collaborative filtering, and predictive models of repeat customer behavior. Such an integration across data science approaches will not only be useful to UMS and arts organizations, but to a broader methodological and customer analytics audience as well. 

    
    Our work demonstrates how data science can help nonprofit organizations further achieve their missions. Data visualizations and business analytics help extract information in underutilized data and identify areas of improvement. The statistical models and machine learning approaches presented here are only a starting point for ways data science can help inform decision making in the performing arts.

\section*{Acknowledgments}
The authors would like to thank UMS, specifically the UMS staff Sara Billman, Anna Prushinskaya, and Mallory Schirr for graciously providing their dataset, as well as working with us to identify important areas of focus for our ongoing collaboration. The authors would also like to thank many other MDST members who indirectly contributed to this project. This work is supported by the National Science Foundation under CAREER grant IIS-1453304, an award that helped facilitate the development of the Michigan Data Science Team. 

\nocite{*}
\bibliographystyle{abbrv}
\bibliography{references}

\end{document}